\documentclass[a4paper,UKenglish,cleveref, autoref, thm-restate]{lipics-v2021}

\usepackage{CJK}
\usepackage{xcolor}

\usepackage{xspace}
\usepackage{float}
\usepackage{makecell}

\newcommand{\FO}{\tools{FO}\xspace}
\newcommand{\FP}{\tools{FP}\xspace}
\newcommand{\FC}{\tools{FC}\xspace}

\newcommand{\tools}[1]{\textsc{#1}}

\newcommand{\painless}{\tools{PaInLeSS}\xspace}
\newcommand{\kissat}{\tools{Kis\-sat}\xspace}

\newcommand{\maple}{\tools{Maple}\xspace}
\newcommand{\mapledl}{\tools{Maple-DL}\xspace}
\newcommand{\maplecomsps}{\tools{MapleCOMSPS}\xspace}
\newcommand{\mapleDL}{\tools{MapleLCMDistChronoBT-DL}\xspace}
\newcommand{\cadical}{\tools{CaDiCaL}\xspace}
\newcommand{\cadicalws}{\tools{CaDiCaLWS}\xspace}

\newcommand{\minisat}{\tools{MiniSat}\xspace}
\newcommand{\cadicalWS}{\tools{CaDiCaL\underline{\;}watch\underline{\;}sat}\xspace}
\newcommand{\kissatMAB}{\tools{Kissat-MAB}\xspace}

\newcommand{\pcomsps}{\tools{P-mcomsps}\xspace}

\newcommand{\pakis}{\tools{PaKis}\xspace}

\title{Revisiting Restarts of CDCL: Should the Search Information be Preserved?} 

\author{Xindi {Zhang}} {
\and \url{https://dezhangxd.github.io} 
}{zhangxd@ios.ac.cn}{https://orcid.org/0000-0001-5541-7194}{}

\author{Zhihan {Chen}} { 
\and  \url{}
}{chenzh@ios.ac.cn}{https://orcid.org/0000-0001-5702-2508}{}

\author{Shaowei {Cai}\footnote{Corresponding author}} {Key Laboratory of System Software (Chinese Academy of Sciences) and State Key Laboratory of Computer Science, Institute of Software, Chinese Academy of Sciences, Beijing, China \and School of Computer Science and Technology, University of Chinese Academy of Sciences, China \and \url{http://lcs.ios.ac.cn/~caisw/} 
}{shaoweicai.cs@ios.ac.cn}{https://orcid.org/0000-0003-1730-6922}{}

\authorrunning{X. Zhang, Z. Chen and S. Cai} 

\Copyright{Xindi Zhang, Zhihan Chen and Shaowei Cai} 

\ccsdesc[500]{Theory of computation~Logic and verification}
\ccsdesc[500]{Theory of computation~Constraint and logic programming}

\keywords{SAT Solving, Cold Restart, Information Forgetting} 

\category{} 

\relatedversion{} 

\supplement{ From the GitHub repository, \url{https://github.com/CDCL-cold-restart/cold-restart}, readers can find the source code and raw data of this paper.}

\nolinenumbers 

\EventEditors{John Q. Open and Joan R. Access}
\EventNoEds{2}
\EventLongTitle{42nd Conference on Very Important Topics (CVIT 2016)}
\EventShortTitle{CVIT 2016}
\EventAcronym{CVIT}
\EventYear{2016}
\EventDate{December 24--27, 2016}
\EventLocation{Little Whinging, United Kingdom}
\EventLogo{}
\SeriesVolume{42}
\ArticleNo{23}

\begin{document}

\maketitle

\begin{abstract}
SAT solvers are indispensable in formal verification for hardware and software with many important applications. 
CDCL is the most widely used framework for modern SAT solvers, and restart is an essential technique of CDCL. 
When restarting, CDCL solvers cancel the current variable assignment while maintaining the branching order, variable phases, and learnt clauses. 
This type of restart is referred to as warm restart in this paper. 
Although different restart policies have been studied, there is no study on whether such information should be kept after restarts. 
This work addresses this question and finds some interesting observations. 

This paper indicates that under this popular warm restart scheme, there is a substantial variation in run-time with different randomized initial orders and phases, which motivates us to forget some learned information periodically to prevent being stuck in a disadvantageous search space. 
We propose a new type of restart called cold restart, which differs from previous restarts by forgetting some of the learned information. 
Experiments show that modern CDCL solvers can benefit from periodically conducting cold restarts. 
Based on the analysis of the cold-restart strategies, we develop a parallel SAT solver. 
Both the sequential and parallel versions of cold restart are more suitable for satisfiable instances, which suggests that existing CDCL heuristics for information management should be revised if one hopes to construct a satisfiable-oriented SAT solver. 
\end{abstract}

\section{Introduction}
The propositional Satisfiability problem (SAT) is a decision procedure that asks for the satisfiability of a given propositional formula.
The propositional formulas are usually presented in Conjunctive Normal Form (CNF), i.e., $F = \bigwedge _i \vee _j \ell_{ij}$.
As the first proved NP-complete problem, SAT is a fundamental problem in computer science~\cite{knuth2015art}.
The Conflict-Driven Clause Learning (CDCL) paradigm~\cite{silva1996conflict} is the most prevalent framework in current SAT solvers.
CDCL is a depth-first searching framework that is equipped with two powerful pruning techniques --- non-chronological backtracking and clause learning. The process resembles a binary search tree, where the two outgoing edges of a node represent the two assignments (true or false) of a variable, and each node along with its paternal path is related to a partial (or full) assignment of the given variable set. 
Thanks to the combination of 
restarting~\cite{DBLP:conf/cp/AudemardS12,biere2015evaluating_restart}, 
branching heuristics~\cite{moskewicz2001chaff,liang2016learning}, 
clause management~\cite{audemard2009predicting,oh2015between},
local search cooperation~\cite{cai2021deep},
and other effective components, CDCL has achieved many remarkable successes in a variety of practical applications, such as electronic design automation (EDA)~\cite{marques2000boolean}, 
Model Checking~\cite{vizel2015boolean},
Satisfiability Modulo Theories (SMT)~\cite{kroening2016decision}, among others.

Restarting contributes significantly to the good performance of CDCL solvers.
Gomes et al. discovered that when solving satisfiable instances, introducing noises into specific heuristics of several complete SAT solvers may result in a run-time variation, which follows a heavy-tailed distribution~\cite{gomes1997heavy}. 
Consequently, they suggested restarting regularly to avoid being stuck in a bad searching direction.
Later, it was found that restarting can also benefit solving unsatisfiable instances, and there are mainly two supportive viewpoints.
On the one hand, restarting can be used to compact the assignment stack and improve the order of decision variables~\cite{DBLP:journals/jsat/HamadiJS09}; On the other hand, there is evidence that restarts assist CDCL in learning higher quality clauses~\cite{DBLP:conf/sat/LiangOMTLG18}.

Restart policies determine when to pause the SAT solver's current searching process and relaunch it from the beginning. 
Several restart policies have been proposed and studied~\cite{biere2015evaluating_restart}.
A famous policy is the so-called \emph{Luby} restart~\cite{DBLP:journals/ipl/LubySZ93,DBLP:conf/ijcai/Huang07}, which restarts according to a static fixed-interval sequence. 
The \emph{Luby} sequence has been proved optimal for a particular class of randomized algorithms called Las Vegas algorithms~\cite{DBLP:journals/ipl/LubySZ93,DBLP:conf/ijcai/Huang07}.
Dynamic restart~\cite{DBLP:conf/cp/AudemardS12}, uses on-the-fly information during the searching process to determine when to restart. For example, Glucose restarts~\cite{DBLP:conf/cp/AudemardS12} when the average Literal Block Distance (LBD)~\cite{audemard2009predicting} of recently learned clauses is larger than the global average LBD. 
Biere and Fr{\"o}hlich (2015) proposed a variant of Glucose's restart strategy based on the \emph{exponential moving averages} (EMA)~\cite{biere2015evaluating_restart}.
Modern CDCL solvers typically incorporate multiple restarting methods, and their frequency of restarts is quite high --- solvers may perform hundreds of restarts every second.


Current popular CDCL restart policies can all be viewed as \emph{warm restarts}, in the sense that they keep the search information and use them after a restart. 
There are mainly three types of information that can be kept after restarts:
the scores of variables are preserved for branching;
promising assignments for variables is preserved for phase saving~\cite{DBLP:conf/sat/PipatsrisawatD07} or rephasing~\cite{DBLP:journals/jair/CaiJAIR22};
the learnt clauses are kept for pruning search space.
Such information is kept and used after restarts, so that the previous computational effort spent would not be wasted~\cite{gomes2009exploiting}. 

However, \textbf{\emph{should these information be preserved after restarts?}} This critical question has not been studied yet. This work aims to answer this question via empirical studies. Our studies lead to some interesting findings, and also new variants of restarts that can bring improvements to state-of-the-art CDCL solvers. 

\paragraph*{Contributions}
This paper shows by experiments that even after decades of improvements, modern CDCL solvers still exhibit significant run-time variation, mainly due to the greedy strategies that re-use previous information after restarts.
The run-time variation suggests that ``\emph{periodically performing thorough restarts, resetting heuristic scores, or discarding learnt clauses, may benefit the performance}''.

To study this research question, we evaluate three cold restart variants and their combinations in CDCL solvers. These restart policies erase some search information and are called {\it cold restarts}. 
The first policy, denoted as FO, \textbf{f}orgets the branching \textbf{o}rder, by using random order instead of preserving the branching scores; 
the second, denoted as FP, \textbf{f}orgets the \textbf{p}hases, by using random phases after restart (note that re-phasing has been studied previously, but our idea uses only random phases); 
finally, the third restart policy \textbf{f}orget all the learnt \textbf{c}lauses after restarting, which is denoted as FC.

We implement our cold restart policies in three representative state-of-the-art CDCL solvers, including
\cadicalWS~\cite{manthey2021cadical},
\mapleDL~\cite{kochemazov2020speeding},
and 
\kissatMAB~\cite{cherif2021combining}.
Generally, our experiments show that by performing such ``colder restart'' periodically, CDCL solvers can be further improved, sometimes significantly. The FO and FP policies are helpful for both satisfiable and unsatisfiable instances. The FC policy helps improve the performance of satisfiable instances but significantly degrades the performance of unsatisfiable instances.
Through the experiments on selecting LBD thresholds for forgetting the learnt clauses during FC, we know: when the objective is to improve the performance on satisfiable instances, we should set the threshold as small as possible.

Additionally, we use the idea of the best cold restart policy, that is, \FO, to improve two state-of-the-art parallel solvers named \pcomsps~\cite{vallade2021new} and \pakis~\cite{tchinda2021hkis}. 
Meanwhile, we apply the parallel cold restart method to \kissatMAB~\cite{cherif2021combining} and the performance is particularly good on the satisfiable instances, and shows promising speedups.
Based on the analysis of the clause forgetting threshold, we adopt a simple clause-sharing strategy that only shares clauses whose LBD$\leq$2, and it produces a solver that is better than state-of-the-art methods on satisfiable instances.
Meanwhile, the idea of \FO can be seen as a reasonable alternative to the configuration selection strategy in parallel portfolios as it is compatible with any CPU core.

Cold restart and its parallel variant work particularly better on satisfiable instances. 
As a result, to develop a satisfiable-oriented SAT solver, it is advisable to rethink and refine certain heuristics, for example, the learned information management policies during restart.

\subparagraph*{Paper Organization}
The rest of this paper is organized as follows:
Section~2 gives basic concepts.
Section~3 shows the heavy-tail phenomenon of modern CDCL.
Section~4 studies three types of ``cold restarts'' and their combinations, and further discusses the role of learnt clauses with different LBD values.
Section~5 presents a parallel implementation based on the idea of forgetting order cold restart.
Section~6 discusses some related works and Section~7 concludes the paper.

\section{Preliminaries}

\subsection{Background Knowledge}

Given a set of Boolean \emph{variable}s $V=\{x_1,x_2,...,x_n\}$, a \emph{literal} is either the positive or negative of a Boolean variable.
A \emph{clause} $C = \bigvee _i x_i$ is a disjunction of literals.
A \emph{Conjunctive Normal Form} (CNF) formula $F = \bigwedge _i C_i$ is a conjunction of clauses.
A complete (partial) \emph{assignment} is a mapping $\alpha: V \rightarrow \{0, 1\}$ that assigns values to all (part of) the variables in $V$.
SAT is the problem of deciding whether there is a complete assignment that satisfies a given formula.

Modern SAT solvers are based on the conflict-driven clause learning (CDCL) approach~\cite{marques1999grasp}, which is a (non-)chronological backtracking search process.
This paper is mainly associated with the branching and restart components of CDCL solvers. 

\textbf{Overview of CDCL Branching}
Modern CDCL branching heuristics aim to pick the variable likely to lead to high-quality conflicts, and the ranking score is usually saved in a heap or a priority queue.
Popular branching heuristics include \emph{Variable State Independent Decaying Sum} (VSIDS)~\cite{moskewicz2001chaff}, \emph{Learning-Rate Based Branching} (LRB)~\cite{liang2016learning} and \emph{Variable Move To Front} (VMTF)~\cite{ryan2004efficient}.
Note that the default order of all the solvers mentioned in this paper is sorted according to the index number of variables.

\textbf{Overview of CDCL Phasing}
The phases of CDCL decide the assignment when branching on a variable. Modern CDCL solvers use the phase saving trick \cite{DBLP:conf/sat/PipatsrisawatD07}, which prefers to assign a variable the value it was last assigned. 
Recently, rephasing \cite{biere2020chasing,DBLP:journals/jair/CaiJAIR22} is proposed to periodically reset the phases. There are studies proposing to stick to phases that maximize the assignment trail by extending promising assignments via a technique called target phases \cite{fleury2020cadical} or via a non-conflicting propagation and improved by local search \cite{cai2021deep}.

\textbf{Overview of CDCL Restarting}
Modern CDCL solvers implement multiple restarting methods, as frequent restarts greatly improve the performance of CDCL.
When restarting, CDCL solvers cancel the partial assignment of all variables, but preserve the previous order, phase, and learned clauses.
The restarting operations of modern SAT solvers are almost the same, but there are various heuristics that decide when to start.
The most implemented restarting heuristic in modern SAT solvers include \emph{Luby}~\cite{DBLP:journals/ipl/LubySZ93}, \emph{the Glucose-style restart}~\cite{audemard2009predicting} and its EMA variants~\cite{biere2015evaluating_restart}. 

\textbf{Overview of Parallel SAT Solving}
There are two main parallel paradigms: the cube-and-conquer paradigm~\cite{heule2011cube}, and the portfolio paradigm. 
The first paradigm divides the search space into millions of little subspaces and solves each of them separately, which is successfully used to solve many mathematical problems~\cite{heule2018schur,heule2016solving}. 
The second paradigm simply runs different SAT solvers or different configurations of a single SAT solver for each thread, where the latter one is called \emph{diversity} \cite{le2017painless,balyo2015hordesat,DBLP:journals/jsat/HamadiJS09,manthey2021mergesat}. 
Modern portfolio solvers usually use clause sharing~\cite{DBLP:journals/jsat/HamadiJS09}
to enhance the performance by avoiding searching the same search space.

\subsection{Experimental Settings}
In this paper, all experiments are carried out on a cluster with two AMD EPYC 7763 CPU @ 2.45Ghz of 128 physical cores in total and 1T RAM under the operating system Ubuntu 20.04 LTS (64bit).
Benchmarks are selected from SAT Competitions 2020 and 2021, SC20 and SC21 for short, and each has 400 instances.
We only perform one test for each solver, as in SAT competitions. The seeds for each solver are 0, and thus those solvers are deterministic.

For each solver and benchmark, we report the number of solved satisfiable (SAT) and unsatisfiable (UNSAT) instances, denoted as `\#SAT' and '\#UNS', and the total solved instances, \#ALL=\#SAT+\#UNS.
We also report the penalized average run time score (PAR2),  which penalizes the run time of a failed run as twice the cutoff time (5000 seconds in default), and is a ranking basis of SAT Competition\footnote{For SAT and UNSAT experiments, the PAR2 time of the solvers is calculated by excluding the instances that none of the solvers (in the competition and in this work) can solve. }.
For parallel solvers, the speedup for $t$ threads is the ratio of the run time of a single core to the run time of $t$ cores.
The parallel solvers in this paper can use at most 32 CPU cores for each instance, which is a common constraint followed by the competitors in SAT Competition. 
The source codes and detailed results of this paper can be found in a GitHub repository~\footnote{\url{https://github.com/CDCL-cold-restart/cold-restart}}.
The base sequential and parallel solvers to implement our methods are as follows:
\begin{itemize}

    \item \textbf{\mapleDL} (Version 3.0) (\mapledl for short) ~\cite{kochemazov2020speeding}: 
    It is a derived version of the \maple series SAT solvers, and these series have won 4 SAT Competitions/Races since \maplecomsps~\cite{liang2016learning}.
    
    \item \textbf{\cadicalWS}  (\cadicalws for short) \cite{manthey2021cadical}: It is developed based on \cadical \cite{fleury2020cadical}. It and its improved version won the \cadical Hack of SC21 and SC22.
    
    \item \textbf{\kissatMAB}~\cite{cherif2021combining}: 
    It is the winner of the Main-Track of SC21, which is based on the SC20 Main-Track winner \kissat \cite{fleury2020cadical}. The main idea is to use MAB to dynamically adjust the VSIDS and CHB branching methods. 

    \item \textbf{\pakis}~\cite{tchinda2021hkis}:
    It is a parallel portfolio based on \kissat and solved the most satisfiable instances in the Parallel-Track of SC21. 
    
    \item \textbf{\pcomsps}~\cite{le2017painless}: \painless series parallel solvers have won the Parallel-Track of SAT Competitions 2018, 2020, and 2021. 
    \pcomsps is a recent portfolio solver based on the \painless framework.
\end{itemize}

\section{Run Time Variation with Warm Restarts}
An intriguing phenomenon of backtrack-style SAT solvers is that their performance on one certain instance can vary dramatically with different random seeds, or when the heuristics change slightly.
Restart policies are proposed to address this heavy-tailed behavior.
Initially, when Gomes et. al. proposed the restart to exploit the run time variation of DPLL solvers (the predecessor of CDCL), they introduced some randomized strategies to the branching orders \cite{gomes1998boosting}. 
Later, in order to re-use previous efforts, more restart policies keep the search information, and we call them warm restarts.
In fact, \minisat keeps branching order scores after restart \cite{een2003extensible}. Nowadays, state-of-the-art CDCL solvers employ warm restarts. In this way, the performance of CDCL solvers can be improved, but the price is that the solvers are sensitive to the heuristics, including the ordering and phasing heuristics.
With warm restart, modern CDCL solvers are inclined to visit the previous search space as they preserve the variable branching scores and phases. 

In this section, we study the run time variation of a representative state-of-the-art CDCL solver w.r.t. the initial setting of branching order and default phases of variables.

\subsection{Investigations on the Initial Order}

\begin{figure}[!t]
\caption{Runtime variations with different initial orders in log scale. The left (right) sub-figure reports the results of SAT (UNSAT) instances with circle (star) markers. The instances are sorted according to \kissatMAB's run time, and we do not report the easy instances that can be solved by \kissatMAB in less than 10 minutes.} \label{fig:ro100}

  \centering
  \includegraphics[width=0.49\textwidth]{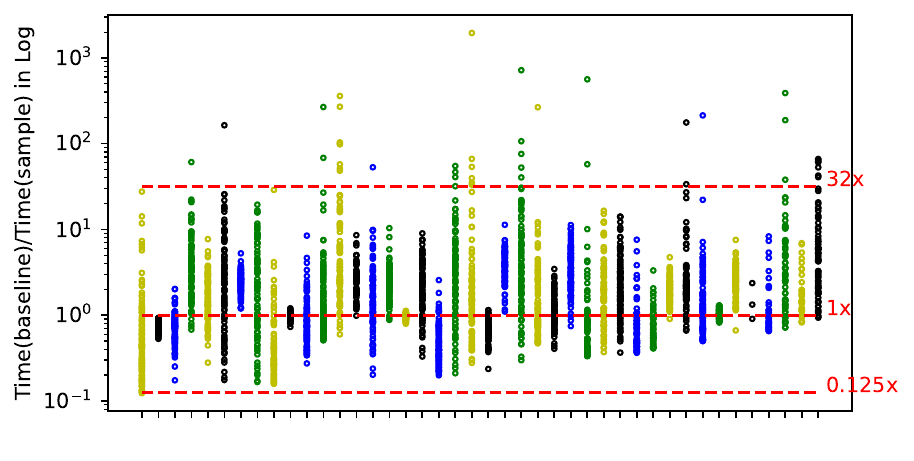}
  \includegraphics[width=0.49\textwidth]{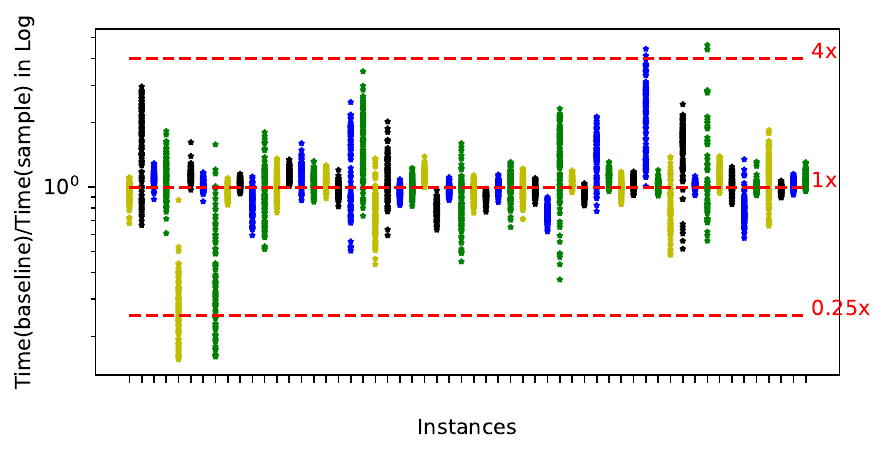}
  
\end{figure}

To study the impact of the initial branching order on the performance of CDCL solvers, we choose \kissatMAB~\cite{cherif2021combining} as the studied solver, the winner of Main Track in SC21. To avoid over-tuning, we use the SC20 benchmark for assessment.
We select the instances that \kissatMAB can solve within 5000s (the cutoff in SAT Competitions), and the run time is denoted as $T_{baseline}$. 

For each selected instance, we run \kissatMAB 100 times with different initial branching orders, which are generated randomly.
The run times are denoted as $T_{sample}$. 
For each instance, Fig.~\ref{fig:ro100} reports the 100 ratios of $\frac{T_{baseline}}{T_{sample}}$, unless $T_{sample}$ is timeout. We filter the easy instances
that \kissatMAB can solve in 10 minutes in this figure.

    For SAT instances, the run time of the solver varies significantly depending on the initial orders. For 33.3\% SAT instances, there are at least one (among the 100) initial orders that can lead to at least $32\times$ speedup. Further, 33 satisfiable instances that cannot be solved by the original solver can be solved by the solver with these random orders.
    
    For UNSAT instances, the run time variance is much smaller. For most UNSAT instances, among the 100 runs with different random orders, the variation for the best run time (or the worst run time) is usually less than 4$\times$.

\subsection{Investigations on the Initial Phases}

\begin{table}[t]
\caption{Results for the random sampling experiment. 
We report the average percent of timeout tries as `Failed', and the average percent of tries that get at least $n\times$ speedup; for the column with $n<1$, the results report the percentage of runs that is at least $n$ times slower than the base solver. }
\label{tbl:pre-op}
\renewcommand
\tabcolsep{12pt}
\centering

\scalebox{0.85}{
\begin{tabular}{|l|l|llll|ll|l|l|l|l|l|l|l|l|}
\hline

 \#Selected & Fail & $2\times$ & $4\times$& $10\times$ & $32\times$ & $0.5\times$ & $0.25\times$ \\
\hline
\multicolumn{8}{|c|}{Random Initial Orders} \\
\hline 
 SAT(151)   &3.58&   25.72& 12.37& 4.72&  1.14&  25.21& 15.03 \\
 UNSAT(121) &2.1&    1.7&   0.06&  0.0&   0.0&   7.78&  5.22\\
 ALL(272)  &2.92&   15.03& 6.89&  2.62&  0.63&  17.46& 10.67\\
\hline

\multicolumn{8}{|c|}{Random Initial Phases} \\
\hline 
 SAT(151)   &3.71&   25.29& 13.23& 3.87&  1.03&  21.95& 10.91 \\
 UNSAT(121) &1.02&   2.95&  0.79&  0.03&  0.0&   5.31&  2.2\\
 ALL(272)  &2.51&   15.35& 7.7&   2.17&  0.57&  14.54& 7.03\\
\hline

\end{tabular}
}

\end{table}

Besides the branching order, the variables' phase is another important part to guide the CDCL search. 
We study the influence of random initial phases on the performance of the solver. The experiments are carried out similarly. 

In Table 1, we summarize the experiment results of the influence of initial orders and initial phases on the run time of the solver. The plot-figure of run time speedup distribution according to phases resembles Figure~\ref{fig:ro100}, so it is not presented here. Similar observations can be obtained as with the study on the random initial branching orders.

We would like to note that recent solvers introduced the rephasing methods to periodically reset the variables' phases. They prefer to pick the phases that can lead to a more depth search space~\cite{fleury2020cadical} or use local search optimized phases~\cite{cai2021deep}. 
Although these rephasing heuristics introduce a small probability of randomizing the phases, 
it is still insufficient to eliminate the heavy-tailed distribution.

The run time variance across different initial phases motivates us to test whether there is a good heuristic that can help to improve performance. 
We try constructing good initial phases by propagation~\cite{DBLP:conf/cp/CaiLZZ21}, which is similar to the `warm-up' ideas of Knuth~\cite{knuth2015art} and how `ReasonLS' initializes the local search~\cite{cai2018reasonls}.
Similarly, for each instance, we generate 100 full assignments using this heuristic method, yet it failed to lead to better peak performance (the best among the 100 runs) than the randomized method.

\subsubsection*{Observation for Warm Restart}

Overall, we observe that the run-time variation is more significant in SAT instances than in UNSAT instances.  For each instance, we calculate the Coefficient of Variation (CV) of the 100 run-times, which is the standard deviation divided by the mean. A higher CV indicates greater dispersion.
For the case of initial branching order (resp. variable phase),  the average CV is 71.5\% (resp. 71.8\%) for SAT instances and 27.2\% (resp. 36.8\%) for UNSAT instances.
We conclude this section by giving this important observation:

\textbf{Observation 1: The performance (run time) of current CDCL solvers is sensitive to the initial branching order and initial phases, particularly for satisfiable instances. }

\begin{table*}[!t]
\caption{Results on the Forgetting Order (FO), Forgetting Phase (FP), and Forgetting Clauses (FC) cold restart policies.}
\label{tbl:FOPC}
\renewcommand\tabcolsep{9.0pt}
\centering
\scalebox{0.8}{
\begin{tabular}{|c|lr|lr|lr|c|}
\hline
Solver  
& \multicolumn{2}{c|}{\small \#SAT \;\;\;\;\;\; PAR2} 
& \multicolumn{2}{c|}{\small \#UNS \;\;\;\;\;\; PAR2}   
& \multicolumn{2}{c|}{\small \#Solved \;\;\;\; PAR2} 
& $p$
\\
\hline 
& \multicolumn{7}{c|}{SAT Competition 2020 (\#400)}  \\
\hline
\mapledl  & 107 & 5043.39 & 113 & 3095.50 & 220 & 5078.55 & - \\
\mapledl+\FO  & 113(+6) & 4846.62 & 111(-2) & 3175.93 & 224(+4) & 5014.36 &  1$\times$E6\\
\mapledl+\FP  & 114(+7) & 4731.01 & 111(-2) & 3179.28 & 225(+5) & 4960.39& 8$\times$E5\\
\mapledl+\FC  & 114(+7) & 4869.95 & 105(-8) &3620.68& 219(-1) & 5190.05 &1$\times$E6 \\
\hline
\cadicalws & 123 & 4247.01 & 120 & 2853.88 & 243 & 4608.88 & -  \\
\cadicalws+\FO & 126(+3) & 3998.46 & 123(+3) & 2704.77 & 249(+6) & 4435.03 & 1$\times$E6 \\
\cadicalws+\FP & 130(+7) & 3891.82 & 122(+2) & 2796.81 & 252(+9) & 4418.16  & 3$\times$E5\\
\cadicalws+\FC & 125(+2) & 4110.95 & 112(-8) & 3298.22 & 237(-6) & 4708.32 & 6$\times$E5\\
\hline
\kissatMAB & 151 & 2510.49 & 121 & 2557.91 & 272 & 3670.18 & - \\
\kissatMAB+\FO & 159(+8) & 2226.81 & 122(+1) & 2515.19 & 281(+9) & 3518.92 & 8$\times$E5\\
\kissatMAB+\FP & 156(+5) & 2363.07 & 122(+1) & 2591.20 & 278(+6) & 3614.02 & 1$\times$E6\\
\kissatMAB+\FC & 158(+7) &2278.36& 109(-12) &3421.98& 267(-5) & 3879.05 & 7$\times$E5\\
\hline
&\multicolumn{7}{c|}{SAT Competition 2021 (\#400)} \\
\hline
\mapledl  & 115 & 3458.14 & 143 & 3076.04 & 258 & 4163.22 & - \\
\mapledl+\FO  & 120(+5) & 3270.75 & 146(+3) & 3067.84 & 266(+8) & 4084.84 & 1$\times$E6\\
\mapledl+\FP  & 118(+3) & 3279.54 & 144(+1) & 3054.42 & 262(+4) & 4082.12 & 1$\times$E6 \\
\mapledl+\FC  & 115 &3520.02& 132(-11) &3637.65& 247(-11) & 4449.51 &1$\times$E6\\
\hline
\cadicalws & 126 & 2667.20 & 144 & 2999.09 & 270 & 3811.46 & -\\
\cadicalws+\FO & 132(+6) & 2483.37 & 144 & 2994.74 & 276(+6) & 3735.90 & 8$\times$E5\\
\cadicalws+\FP & 132(+6) & 2501.37 & 143(-1) & 3049.62 & 275(+5) & 3768.62 & 3$\times$E5 \\
\cadicalws+\FC & 130(+4) & 2557.67 & 138(-6) & 3251.08 & 268(-2) & 3884.82 & 6$\times$E5 \\
\hline
\kissatMAB & 142 & 1654.57 & 147 & 2714.55 & 289 & 3274.09 & - \\
\kissatMAB+\FO & 143(+1) & 1624.87 & 152(+5) & 2555.97 & 295(+6) & 3188.47 & 4$\times$E5 \\
\kissatMAB+\FP & 142 & 1655.81 & 150(+3) &2634.91& 292(+3) & 3237.56 & 1$\times$E6 \\
\kissatMAB+\FC & 143(+1) &1677.60& 137(-10) &3339.01& 280(-9) & 3573.68 & 7$\times$E5\\
\hline
\end{tabular}
}
\end{table*}

\section{Evaluating Cold Restarts} 

For many combinatorial search algorithms, random noises could cause run time variance, and restart is an effective method to lessen this variance~\cite{baptista2000using,gomes1998boosting}.
However, modern CDCL solvers often employ frequent restart policies, and only a few conflicts occur between two successive restarts. Therefore, modern CDCL solvers tend to search in a similar branching order as the one before restart, since the variable branching scores are adjusted primarily depending on conflicts~\cite{DBLP:conf/sat/RamosTH11}.
Meanwhile, modern SAT solvers tend to assign the same truth value for a chosen variable due to the usage of phase saving~\cite{DBLP:conf/sat/PipatsrisawatD07}. Consequently, the current CDCL restart method directs the search direction after a restart to a search space near the location before the restart.


According to the findings of Section~3, modern solvers still exhibit a significant run time variance according to different initial branching orders or phases, which may be caused by the information-saving behavior of restart policies.
In order to reduce the probability of getting stuck in a disadvantageous searching space, in this section, we study the central research question in this work:

\textbf{Should the search information be kept after restart?}

We refer to the restart policies saving all search information in between restarts as \emph{warm restarts}. All modern CDCL solvers adopt warm restarts. 
There are mainly three types of information that are kept in the warm restart, including branching order (variable score),  phase (variable polarity), and learnt clauses.
For convenience, we refer to the restarts forgetting all search information as \emph{complete cold restarts}, and those forgetting parts of the information as \emph{partial cold restarts}. Both of them belong to \emph{cold restarts}, distinguishing from the normal warm restarts in modern CDCL solvers.

In this section, we first study three cold restart policies (Table~\ref{tbl:FOPC}), and then study their combinations (Table~\ref{tbl:hybrid}).
\begin{itemize}
    \item FO policy: Cold restart that forgets the branching order
    \item FP policy: Cold restart that forgets the phases
    \item FC policy: Cold restart that forgets learnt clauses
\end{itemize}


Instead of replacing the warm restarts in the CDCL solvers with our cold restarts, we indeed integrate the cold restarts into the solvers while keeping the warm restarts.\footnote{Experiments indicate that the performance deteriorates significantly when warm restarts are directly replaced with cold restarts because cold restarts with an excessively high frequency severely disrupt the search process. It can even make the solver incomplete.} 
In essence, the cooperation of warm restarts and cold restarts corresponds to the classic topic in search algorithms: the balance between exploitation (warm restarts) and exploration (cold restarts).

\textbf{ Cold Restart Frequency} 
We use a common static restart policy with linear growth intervals to keep completeness.
Let us use $r$ to denote the number of conflicts encountered since the last cold restart, and $n$ the number of cold restarts performed.  When $r \geq p\times n$ the next warm restart is replaced by the cold restart, where $p$ is a parameter.
We use the same interval metric (the number of conflicts) as most current restart policies, which keeps the number of cold restarts positively correlated to the number of warm restarts.\footnote{The number of cold restart is usually fewer than a dozen for most instances and the median number is 3 for \kissat+FO, while the number of warm-restart is approximately 3, 4 orders of magnitude more than that.}

\textbf{Parameter Tuning} There is only one hyper-parameter $p$ for the cold restart intervals. For each benchmark and solver, the parameter $p$ is adjusted from 100k conflicts to 1000k conflicts with 100k as the minimum granularity, which results in 10 numbers. We report the best results among the 10 parameters. The specified settings for each solver are given in the `$p$' column for each table, and all the numbers are given in scientific notation.  Note that we did not tune $p$ separately for satisfiable and unsatisfiable benchmarks. 
\subsection{Forgetting the Branching Order}
Each time the FO (Forgetting Order) cold restart is performed, for all variables, the branching score for each branching heuristics is reset to a float number in [0, 1] randomly, 
because it is a relatively small number, whose impact on the scoring of other variables can be ignored after a few steps of scoring.
The data structures used to sort the variables are updated accordingly.
Note that for the heuristic with no variable scores, e.g. VMTF, we shuffle the variable order that enters the priority queue.

From the results, we can see that periodically resetting variable orders randomly improves the overall performance, especially for SAT instances. In detail, by implementing order forgetting, the \#Solved number is increased by 6.3 and 6.7 on average for each solver on SC20 and SC21, respectively. The number for \#SAT are 5.7 and 4.0. The improvement for the SAT instances mainly owes to the huge run time variance depending on different initial branching orders.

We also observe the improvements on \#UNS for all the solvers, particularly for \kissatMAB.
A reasonable explanation is that restart helps to find better refutation paths with high-quality clauses \cite{DBLP:journals/jsat/HamadiJS09}. 

\textbf{Observation 2:  The cold restart that forgets the branching order information is helpful for solving both SAT and UNSAT instances.}

\subsection{Forgetting the Phases}
Now we turn to the FP (Forgetting the Phases) cold restart policy.
Each time the FP restart is performed, we assign a random value either 0 or 1 for each variable.
Similarly, we implement the method on top of the base  solvers. 

The FP cold restart allows to improve the overall performance of all the CDCL solvers.
The number of \#Solved is improved by 4.5, 7.0, and 4.5 on average for the three solvers (in the order of their appearance in the table) for each benchmark. We also observe that the performance improvements are mainly due to the improvements on solving satisfiable instances, which is similar to FO. 

\textbf{Observation 3: The cold restart that forgets the phases is helpful for solving SAT instances, while it has a slight impact on the performance of UNSAT instances.}

\subsection{Forgetting Learnt Clauses}

The FC (Forgetting Learnt Clauses) cold restart policy is simple: Each time a FC cold restart is executed, the solver deletes all the learnt clauses that are maintained by the internal data structures related to clauses management.
We implement FC restart in the three base solvers.

We can observe that forgetting all the learnt clauses is always harmful to solving UNSAT instances. Nevertheless, it sometimes can be beneficial for solving satisfiable instances. Indeed, it is helpful for all the tested solvers on solving SAT instances.
Particularly, the FC restart helps to solve 7 more satisfiable instances for both  \mapleDL and \kissatMAB on SC20. This is somehow surprising.
Modern CDCL solvers have a clause management component, which periodically removes some bad clauses to release the reasoning burden. Our experiments demonstrate periodically deleting all learnt clauses can be helpful for solving satisfiable instances, which motivates further study on the role of learnt clauses, particularly in the context of solving a satisfiable instance.

\subsubsection{LBD Threshold for Forgetting Learnt Clauses}
In order to gain a better understanding of the role that learned clauses play in CDCL,
we carried out additional supplementary experiments for FC. 
Based on these experiments, we can learn which clauses should be forgotten during cold restart.
We established different thresholds for forgetting learnt clauses during FC cold restart.
In the study that proposed the famous three tier clauses management strategy, Chanseok Oh stated that the learnt clauses with LBD>5 are barely useful in a practical and global sense~\cite{oh2015between}.
Thus, we set the threshold from LBD > 0 (the default version of FC cold restart) to LBD > 5, in order to preserve some useful learnt clauses~\footnote{
When a learned clause is produced, if its LBD is 1, it signifies a unit clause, which is immediately used in assigning a value to a variable. Consequently, such clauses are not stored in the learned clause database. Therefore, there will be no learned clauses with LBD less than 2 in the database if there is no other reason. However, it is possible to encounter learned clauses with LBD equal to 1 when performing FO. This occurs when these clauses undergo simplification or their LBDs are recalculated under another branching order when used.}.
We take \kissatMAB as the base solver and use benchmarks from SAT Competition 2020 and 2021 for evaluation. Note that the cold-restart parameter $p$ is the same as the corresponding default FC version. 
From the results in Table~\ref{tbl:FC_lbd}, we learn that:
\begin{itemize}
    \item Forgetting more learning clauses usually means better satisfiability performance but worse unsatisfiability performance.
    \item Keep clauses with LBD $\leq 3$ during cold restart helps the solver reach the most balanced performance, which provides another evidence for Oh's methods that take LBD=3 as the cut between core learnt clauses and the mid-tier learnt clauses~\cite{oh2015between}.
\end{itemize}

% ...

\begin{table}[!tbp]
\caption{FC versions that forget learnt clauses according to different LBD thresholds. }
\label{tbl:FC_lbd}
\centering
\renewcommand\tabcolsep{10.0pt}
\scalebox{0.8}{
\begin{tabular}{|l|lr|lr|lr|}
\hline 
Forgetting Type &\#SAT & PAR2 &\#UNS & PAR2 &\#Solved & PAR2 \\
\hline
\multicolumn{7}{|c|}{SAT Competition 2020 (\#400)} \\
\hline
LBD > 0  & 158 & 2278.36 & 109  & 3421.98 & 267 & 3879.05 \\
LBD > 1  & 155 & 2385.65 & 115 & 3082.32 & 270 & 2689.80 \\
LBD > 2  & 153 & 2428.49 & 120 & 2622.74 & 273 & 2513.30 \\
LBD > 3  & 154 & 2405.37 & 120 & 2578.99 & 274 & 2481.17 \\
LBD > 4  & 153 & 2447.50 & 122 & 2494.85 & 275 & 2668.17 \\
LBD > 5  & 153 & 2495.89 & 121 & 2542.93 & 274 & 2516.38 \\

\hline

\multicolumn{7}{|c|}{SAT Competition 2021 (\#400)} \\
\hline
LBD > 0  & 143 & 1677.60 & 137 &3339.01 & 280 & 3573.68 \\
LBD > 1  & 144 & 1626.29 & 141 & 3072.62 & 285 & 2403.80 \\
LBD > 2  & 142 & 1708.98 & 142 & 2924.58 & 284 & 2362.45 \\
LBD > 3  & 142 & 1742.83 & 151 & 2636.28 & 293 & 2223.13 \\
LBD > 4  & 140 & 1748.33 & 149 & 2669.04 & 289 & 2243.28 \\
LBD > 5  & 138 & 1840.09 & 150 & 2643.33 & 288 & 2271.89 \\
\hline
\end{tabular}
}
\end{table}

\subsubsection{Learnt Clause Utilization}

We have also calculated the utilization efficiency of clauses belonging to different LBD values in the CDCL process. 
Specifically, for LBD values ranging from 2 to 7, we have separately tracked the usage of learned clauses at these levels in both unit propagation and conflict analysis.
A learned clause is considered to be involved in conflict analysis once if it participates in the generation of another learned clause, and it is deemed to be involved in unit propagation once if it contributes to the reasoning of a variable's assignment directly. For the convenience of statistics, the LBD level of a learnt clause is considered to be the LBD value calculated when it was generated.

The data are shown in Table~\ref{tbl:fc_lbd_usage}, and we normalized the statistics by dividing them by the corresponding values in the `LBD=2' column. From the results, we learnt that there is a significantly larger disparity between the data in columns `LBD=2' and `LBD=3', while the differences among the other columns are relatively minor.

\begin{table}[!tbp]
\caption{Normalized statistics of the usage of learning clauses with different LBD values.}
\label{tbl:fc_lbd_usage}
\centering
\renewcommand\tabcolsep{10.0pt}
\scalebox{0.8}{
\begin{tabular}{|l|l|l|l|l|l|l|}
\hline 
Type & LBD=2 & LBD=3 & LBD=4 & LBD=5 & LBD=6 & LBD=7 \\
\hline
conflict & 1.0 & 0.2641 & 0.1887 & 0.1381 &  0.1004 & 0.0670 \\
propagate & 1.0 & 0.1899 & 0.1289 & 0.0833 & 0.0447 &  0.0215\\
\hline
\end{tabular}
}
\end{table}


\textbf{Observation 4:  The cold restart that forgets all learnt clauses is always harmful to solving UNSAT instances, while it is usually (not always) beneficial for solving satisfiable instances. Keeping learnt clauses within higher LBD in cold restart helps for the UNSAT instances.}

\subsection{Forgetting Multiple Types of Information}

In the preceding section, we evaluated three cold restarts that forget only one type of information, including orders, phases, and learnt clauses.
In this subsection, we evaluate the performance of forgetting multiple types of information, leading to four combinations, including, `+FO+FP', `+FO+FC', `+FP+FC', and `+FO+FO+FC'.

\begin{itemize}
    \item \FO+\FP: Each  time this cold restart is executed, the solver shuffles the branching order and resets the phase randomly.
    \item \FO+\FC: Each time this cold restart is executed, the solver randomly shuffles the branching order and removes all learnt clauses.
    \item \FP+\FC: Each time this cold restart is executed, the solver resets the variable phases randomly and removes all learnt clauses.
    \item \FO+\FP+\FC: Each time this cold restart is executed, the solver shuffles the branching order randomly, resets the variable phases randomly, and removes all learnt clauses.
\end{itemize}

We report the best version of hybrid cold restart for each solver on each benchmark in Table \ref{tbl:hybrid}. The results indicate that there is likely optimal configuration for cold restarts for SAT and UNSAT instances respectively.

\textbf{Observation 5:  The FO+FC cold restart gives the best performance for satisfiable instances. The FO restart is the best for UNSAT instances and also gives the overall (SAT+UNSAT) best performance in the most cases.}

\begin{table*}[!t]
\caption{
For each solver, we report the best configuration (`BConf') of hybrid cold restart and the corresponding number of changed solved SAT (resp. UNSAT, all) instances in the $\Delta_{\mathrm{SAT}}$ (resp. $\Delta_{\mathrm{UNS}}$, $\Delta_{\mathrm{ALL}}$) columns, compared to the original solver. 
The best configuration is the solver that solves the most instances, breaking ties in favor of the minimal PAR2 score.
}
\label{tbl:hybrid}
\centering
\renewcommand\tabcolsep{9pt}{
\scalebox{0.8}{
\begin{tabular}{|c|lc|c|lc|c|lc|c|}
\hline
Solver  
& \multicolumn{2}{c|}{\small BConf \;\;\;\; $\Delta_{\mathrm{SAT}}$} &$p$ 
& \multicolumn{2}{c|}{\small BConf \; $\Delta_{\mathrm{UNS}}$} &$p$  
& \multicolumn{2}{c|}{\small BConf \; $\Delta_{\mathrm{ALL}}$} &$p$ 
\\
\hline
& \multicolumn{9}{c|}{SAT Competition 2020 (\#400)} \\
\hline
\mapledl   & +FO+FC & +8  &1$\times$E6 & +FO+FP & -1  &4$\times$E5 & +FP & +5 &8$\times$E5 \\
\cadicalws & +FO+FC & +12 &8$\times$E5 & +FO & +3     &1$\times$E6 & +FP & +9 &3$\times$E5 \\ 
\kissatMAB & +FO+FC & +10 &4$\times$E5 & +FO & +1     &8$\times$E5 & +FO & +9 &8$\times$E5 \\
\hline

& \multicolumn{9}{c|}{SAT Competition 2021 (\#400)} \\
\hline
\mapledl   & +FO & +5    &1$\times$E6 & +FO & +3    &1$\times$E6 & +FO & +8 &1$\times$E6\\
\cadicalws & +FO+FC & +6 &1$\times$E6 & +FO+FP & +1 &1$\times$E6 & +FO & +6 &8$\times$E5\\ 
\kissatMAB & +FO+FC & +3 &4$\times$E5 & +FO & +5    &4$\times$E5 & +FO & +6 &4$\times$E5\\
\hline
\end{tabular}
}
}
\end{table*}

\subsection{Detailed Analysis for Preference}
From the previous results, we know the FO cold restart shows the best overall performance. This motivates us to analyze the preference of the FO cold restart on different types of instances.
We split the benchmarks into small categories according to the proceedings of SC20 and SC21. 
We summarized the results here, and the detailed results for all categories are given in our GitHub repository to save space. `\#$k$' is used to denote the number of instances belonging to a given family. 

\begin{itemize}
    \item Our method is more suitable for solving the `hypertree decomposition' family (\#14), `sliding tile puzzle' family (\#13), some traveling salesman problems like the `minimal super-permutation' family (\#13), and some instances belong to the coloring problem. Compared to the base solver, the \FO improved solver can solve at least two more instances for each of these families.
    \item For some families, the base solver, and the +\FO version show complementary performance. 
    For example, \kissatMAB and \kissatMAB+\FO both solve 8 instances on the `circuit multiplier' family (\#13), but there are four complementary examples. This shows a potential for further improvements by a deeper study on the balance of warm restart and cold restart.
    \item On the other hand,
    the FO method deteriorates the performance on the `preimage' family (\#11), which encodes the preimage attack problem from cryptanalysis. \kissatMAB+\FO solves 2 fewer instances in this family.
    \item 
    The CV value of solvers on SAT instances remains high, exceeding 50\% for half of the instances, slightly improved compared to the version utilizing solely warm-restarts. The `baseball lineup selection' family (\#14) exhibits a CV value of less than 30\%.
\end{itemize}

\section{A Parallel Method based on Forgetting Order}

We notice that the cold restart augmented solvers and their original base solver exhibit complementary performance for certain instances. 
This is because a cold restart may interrupt the search process when CDCL fails to find a result in a hard but good search space within the given time.
It drives us to develop a parallel variant of the cold restart, to make better use of the complementarity. 
This section implements the parallel forgetting order (FO) cold restart technique, as it is the version with the best overall performance, and we use it to improve state-of-the-art parallel solvers.

\begin{table}[t]
\caption{Experiment results of parallel forgetting order on two SOTA parallel solvers. The suffix (p) denotes the solver is implemented in parallel. The cold restart parameter $p$ for \pcomsps+\FO(p) and \pakis+\FO(p) are 6$\times$E5 and 4$\times$E5 respectively.}
\label{tbl:FO-parallel}
\centering
\renewcommand\tabcolsep{8.0pt}
\scalebox{0.85}{

\begin{tabular}{|l|lr|lr|lr|}
\hline 
Solver  
&\#SAT & PAR2
&\#UNS & PAR2
&\#Solved & PAR2
\\
\hline
\multicolumn{7}{|c|}{SAT Competition 2020 (\#400)} \\
\hline
\pcomsps(p)      & 158 & 2169.01 & 140 & 843.34 & 298 & 2872.74 \\
\pcomsps+\FO(p)  & 160(+2) & 2112.08 & 141(+1) & 813.09 & 301(+3) & 2834.36 \\
\hline
\pakis(p)        & 176 & 1005.26 & 130 & 1801.21 & 306 & 2671.46 \\
\pakis+\FO(p)    & 181(+5) & 800.08  & 130 & 1776.44 & 311(+5) & 2564.32 \\

\hline

\multicolumn{7}{|c|}{SAT Competition 2021 (\#400)} \\
\hline
\pcomsps(p)      & 143 & 1332.89 & 179 & 725.23 & 322 & 2220.39 \\
\pcomsps+\FO(p)  & 149(+6) & 1002.65 & 178(-1) & 778.81 & 327(+5) & 2113.21 \\
\hline
\pakis(p)        & 155 & 472.05 & 164 & 1705.67 & 319 & 2331.96 \\
\pakis+\FO(p)    & 158(+3) & 173.68 & 166(+2) & 1678.34 & 324(+5) & 2249.03 \\

\hline
\end{tabular}
}

\end{table}

\subparagraph*{Implementation}
Parallel FO uses the same policy as the sequential version except that parallel FO uses different seeds for each thread.

\textbf{Evaluations} 
We implement the parallel \FO cold restart on top of two SOTA parallel solvers, namely \pakis\footnote{We change the base sequential CDCL solver of \pakis from \kissat~\cite{fleury2020cadical} to a better version \kissatMAB~\cite{cherif2021combining} for comparison, and the performance is almost the same with the original one.}
and \pcomsps.
From the results in Table~\ref{tbl:FO-parallel}, FO can be used to improve the performance of them, especially for $\#$SAT. Specifically, \pcomsps+FO(p) (resp. \pakis+FO(p)) improves the PAR2(SAT) time of the \pcomsps (resp. \pakis) by $13.7\%$ ($41.81\%$) on average.



\subparagraph*{Speedup Analysis}
For better evaluating the average speedup\footnote{Only instances that can be solved by both a single core and C cores are included in the calculation of the average speedup of C cores, but the instances should not be too easy (can be solved by both solvers in 1 second). Another proper method for calculating averaged speedup, which considers the timeout instances and the punishment, is to compare the PAR2 scores.} 
of parallel forgetting order, we develop a parallel SAT solver with FO, \kissatMAB+FO(p), which is developed directly upon a sequential SAT solver \kissatMAB without any other parallel techniques.
The results are summarized in Table~\ref{tbl:speedup}.
The solved number and speedup are positively related to the number of CPU cores. The performance on SAT instances is particularly good and it outperforms the SOTA solvers under the same CPU cores, while the improvement on UNSAT instances is limited.

\textbf{Clause selection in clause sharing portfolios}
Prevalent parallel SAT solvers usually employ clause sharing~\cite{DBLP:journals/jsat/HamadiJS09} to exchange learnt clauses among each thread. 
The idea of sharing the learnt clauses among threads is similar to the idea from the current warm restart that keeps learnt clauses of good quality.
This raises our interest in studying whether the rules about the threshold of learning clauses in FC are also valid for parallel clause sharing.

According to the results in Table~\ref{tbl:fc_lbd_usage}, LBD$\leq$3 learnt clauses have the highest cost-effectiveness. 
We extend the basic parallel FO with clause sharing by simply sharing the learnt clause whose LBD$\leq2$ and LBD$\leq3$ with the help of the \painless framework~\footnote{We also tried the clause sharing strategy in \painless, which dynamically adjusts the threshold according to the shared literals. It has a similar performance to the version that shares clauses of LBD$\leq2$, and can draw similar conclusions.}.
For every thread, a maximum of 1500 literals can be shared within a duration of 0.75 seconds.

According to the results and speedups in Table~\ref{tbl:speedup}, clause sharing can further enhance the performance of UNSAT instances, but has almost little effect on SAT instances.
Thus, the parallel FO is good enough to develop a satisfiable-oriented parallel solver.

\vspace{1em}
\textbf{Observation 6: The FO cold restart is beneficial for parallel solving, which is significant for SAT instances and minor for UNSAT instances. Clause Sharing has a significantly positive influence on the UNSAT instances, while has nearly no influence on the SAT instances~\footnote{The observation of the clause sharing behavior is similar to the idea from Balyo et al\cite{balyo2015hordesat}. }.}


\begin{table}[t]

\caption{
Results of parallel solvers with different CPU cores (C) ranging from 1 to 64. We select the SC20 benchmark for testing, and report the solved number, PAR2, and average speedup for SAT, UNSAT, and ALL instances.
}
\label{tbl:speedup}
\centering
\scalebox{0.8}{
\renewcommand\tabcolsep{15.0pt}
\begin{tabular}{|l|lll lll|}

\hline
C 
&{\makecell[c]{\scriptsize \textbf{\#SAT}\scriptsize(\textbf{PAR2})}}  
&{\makecell[c]{\scriptsize \#UNS\scriptsize(PAR2)}} 
&{\makecell[c]{\scriptsize \#Solved\scriptsize(PAR2)}} 
&\multicolumn{3}{c|}{\makecell[c]{\scriptsize Average Speedup\\ \scriptsize \textbf{SAT} \;\;\;\;\;\;\;\; UNS \;\;\;\;\;\;\;\; ALL}} \\
\hline 
\multicolumn{7}{|c|}{\kissatMAB+FO(p), without clause sharing} \\
\hline

1  & 151(2510) & 121(2558) & 272(3670) & 1.0 & 1.0 & 1.0\\ 
2  & 161(1986) & 123(2440) & 284(3376) & 4.2 & 1.13 &  2.8 \\
4  & 168(1477) & 123(2404) & 291(3120) & 12.5 & 1.16  & 7.5 \\
8  & 176(1190) & 125(2317) & 301(2951) & 10.0  & 1.18  & 6.1 \\
16 & 177(1024) & 124(2320) & 301(2872) & 15.0  & 1.22 & 8.9 \\
32 & 182(766)  & 125(2209) & 307(2707) & 26.8  & 1.32 & 15.5 \\
64 & 183(667)  & 125(2231) & 308(2668) & 27.5  & 1.35 & 16.0 \\

\hline
\multicolumn{7}{|c|}{\kissatMAB+FO(p) + \textbf{Sharing (LBD\textbf{$\leq$}2)}} \\
\hline 

1  & 
     151(2510) & 121(2558) & 272(3670) & 1.0 & 1.0 & 1.0 \\
2  & 
     162(1958) & 126(2159) & 288(3259) & 3.5 & 1.3 & 2.5\\
4  & 
     167(1618) & 127(1984) & 294(3032) & 7.5  & 1.9 & 5.0\\
8  & 
     171(1329) & 129(1723) & 300(2797) & 9.4  & 2.9 & 6.5\\
16 & 
     178(924)  & 133(1436) & 311(2498) & 17.0 & 4.6 & 11.5\\
32 & 
     180(770)  & 131(1493) & 311(2445) & 22.1 & 6.6 & 15.4\\
64 & 
     184(583)  & 133(1354) & 317(2304) & 29.0 & 9.4 & 20.5\\

\hline

\multicolumn{7}{|c|}{\kissatMAB+FO(p) + \textbf{Sharing (LBD$\leq$3)}} \\
\hline 

1  &  151(2510) & 121(2558) & 272(3670) & 1.0 & 1.0 & 1.0 \\
2  &  163(1859) & 128(1987) & 291(3148) & 2.6 & 1.6 & 2.2\\
4  &  167(1603) & 130(1785) & 297(2951) & 5.8  & 2.2 & 4.2\\
8  &  173(1197) & 131(1556) & 304(2672) & 10.2  & 3.4 & 7.2\\
16 &  178(911)  & 134(1316) & 312(2447) & 20.7 & 5.2 & 13.9\\
32 &  177(948)  & 135(1169) & 312(2410) & 20.0 & 8.0 & 14.7\\
64 &  182(662)  & 140(891) & 322(2171) & 35.5 & 11.0 & 24.8\\

\hline
\end{tabular}
}
\end{table}

\textbf{Discussions on Comparing Parallel \FO and Previous Diversity Parallel Methods}
In a sense, the cold restart parallel method can be seen as a type of the diversity-based parallel method.
As with previous diversity methods, our parallel FO method utilizes the complementarity between threads. 
Previous methods usually make use of the complementarity between different configurations for the strategies in the CDCL solvers \cite{le2017painless,balyo2015hordesat,DBLP:journals/jsat/HamadiJS09,manthey2021mergesat,tchinda2021hkis}.
For the parallel FO method, the complementarity mainly comes from periodically switching to different starting points of the search space.
Parallel FO is lightweight and easy to implement --- the developers do not need to be familiar with the strategies/codes of CDCL in detail to select configurations.
Simply applying the relevant parallel FO framework with clause sharing can achieve the performance of mainstream parallel SAT solvers.
Another advantage of parallel FO is its strong adaptability. It has good scalability, as it can be applied to any given number of CPU cores without changing the codes.
Based on our experimental findings, it appears that the speedup possesses the potential to further grow alongside an increase in the number of cores.

\begin{figure}[tp!]
\caption[l]{
Cumulative Distribution Function (CDF) for comparing our best sequential and parallel version with the best sequential and parallel competitor for the SAT and UNSAT instances of SC20 and SC21. The x-axis is the runtime in seconds, the y-axis is the number of solved instances. The higher the curve, the better the corresponding solver. 
}\label{fig:cdf}

\centering

  \subfloat{
    \includegraphics[width=0.35\textwidth]{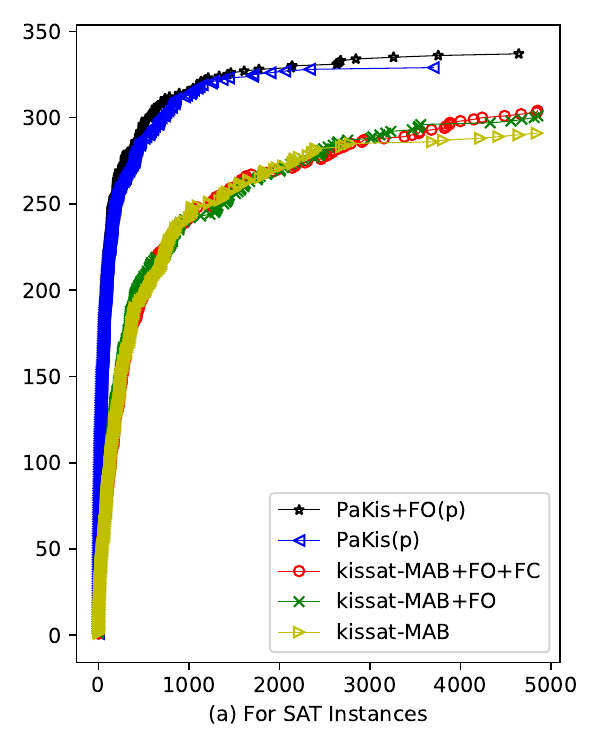}
  }
  \subfloat{
    \includegraphics[width=0.35\textwidth]{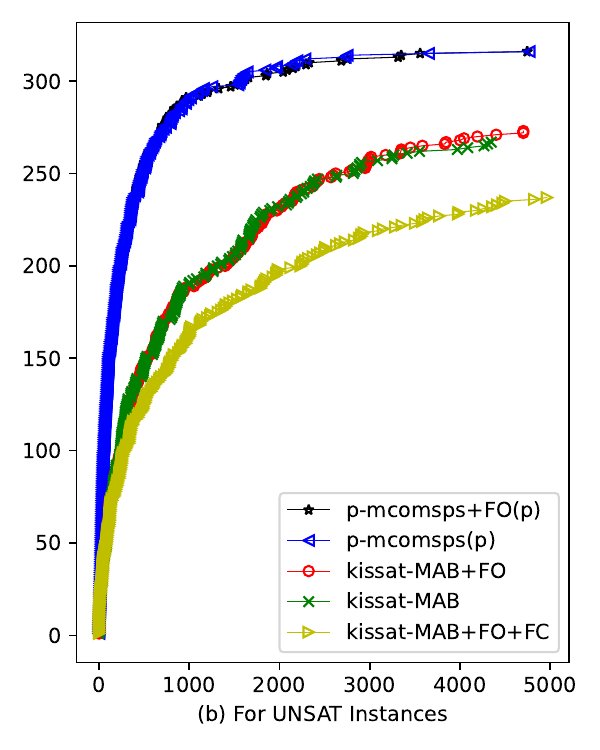}
  }
  
  
\end{figure}

\subparagraph*{Effectiveness Summary}
To see how our methods push the state-of-the-art, we compare our best sequential and parallel solvers with their best-performing competitors in Figure~\ref{fig:cdf}. As we can see from the figure, both the sequential and the parallel versions of cold restart are mainly helpful to push the edge of CDCL solvers \textbf{on satisfiable instances}.

For sequential solvers, \kissatMAB+FO+FC performs the best on SAT instances but deteriorates on UNSAT instances, while \kissatMAB+FO has the best UNSAT performance (Observation 5).
For the parallel version, the parallel FO helps to improve the effectiveness and efficiency of the satisfiable performance, while having a slight influence on the unsatisfiable performance (Observation 6).

\section{Related Works}

Gomes et al. introduced randomization to combinatorial backtrack search and use restarts to exploit the heavy-tailed phenomena~\cite{gomes1997heavy}. 
They introduced controlled randomization into DPLL algorithms by random tie-breaking methods and used a restart heuristic with linear increasing intervals in the following year~\cite{gomes1998boosting}. The CDCL solver Chaff~\cite{moskewicz2001chaff} proposed the famous VSIDS branching heuristic and added a certain amount of transient randomness to the decision procedure, but it kept the current order at restart. Different from the above works, in this work, the FO cold restart randomly shuffled the branching order. 

After the clause learning technique was invented, there was a study on GRASP~\cite{marques1999grasp} showing that recording clauses in between restarts can reduce the run time \cite{baptista2000using}. This quickly became a standard of CDCL solvers.  To the best of our knowledge, no CDCL solver is forgetting all learnt clauses in the search process. 

There have been many works studying when to restart~\cite{een2003extensible,biere2015evaluating_restart,audemard2009predicting,haim2009restart,DBLP:conf/sat/LiangOMTLG18}. However, all these CDCL solvers implemented warm restarts. 
Recently, Biere et al. showed that scrambling CNF usually has a negative effect on the performance of single solvers but has no impact on solver ranking~\cite{biere2019effect}.
Rephasing~\cite{biere2020chasing} is a technique to reset the phases heuristically when restarting, which is becoming a standard technique for current SAT solvers~\cite{fleury2020cadical,cai2021deep}.

\section{Conclusions and Future Work}
Modern CDCL solvers adopt the warm restart policies, which keep all search information in between restarts.
This paper empirically studied the question of whether this information should be preserved in between restarts, leading to several meaningful observations, as well as several cold restart policies (forgetting some information in between cold restarts).  
Experiments suggest that periodically performing cold restarts helps both sequential and parallel CDCL solvers, especially for satisfiable instances.

This work provided an interesting direction for CDCL solvers. In the future, we intend to design better cold restart policies with dynamic intervals.
On real-world benchmarks, even with integrated cold restarts, current CDCL solvers still have great runtime variations, which need to be improved experimentally and theoretically in the future.
Meanwhile, the paper can be viewed as a successful attempt at developing a satisfiable-oriented SAT solver by revisiting the restart policies.

\newpage

\bibliography{main}

\end{document}